# Educational Technology as Seen Through the Eyes of the Readers

Peter Kraker

Know-Center
Inffeldgasse 13/VI, 8010 Graz
E-Mail: pkraker@know-center.at

**Abstract:** In this paper, I present the evaluation of a novel knowledge domain visualization of educational technology. The interactive visualization is based on readership patterns in the online reference management system Mendeley. It comprises of 13 topic areas, spanning psychological, pedagogical, and methodological foundations, learning methods and technologies, and social and technological developments.

The visualization was evaluated with (1) a qualitative comparison to knowledge domain visualizations based on citations, and (2) expert interviews. The results show that the co-readership visualization is a recent representation of pedagogical and psychological research in educational technology. Furthermore, the co-readership analysis covers more areas than comparable visualizations based on co-citation patterns. Areas related to computer science, however, are missing from the co-readership visualization and more research is needed to explore the interpretations of size and placement of research areas on the map.

**Biographical notes:** Peter Kraker is a postdoc at Know-Center, Graz University of Technology. His main research interests are scholarly communication on the web, Science 2.0, and alternative metrics for science and research (altmetrics). He received his doctoral degree (Dr. rer. soc. oec.) from University of Graz in 2014. Peter has worked in a number of EU-funded projects studying online scholarly communication in educational technology, including the Network of Excellence STELLAR and the Marie Curie project TEAM. For his work, he was awarded a Panton Fellowship of the Open Knowledge Foundation and a Marshallplan Scholarship of the Austrian Marshallplan Foundation.

Forthcoming article in the International Journal of Technology Enhanced Learning.

# 1 Introduction

Educational technology is an ever-changing research field. It is influenced by novel pedagogical concepts and emerging technologies (Siemens and Tittenberger 2009), as well as social change (Czerniewicz 2010). Educational technology is also a multi-disciplinary field; education, computer science, and psychology are the three major contributing disciplines. As a result of this multitude of influences, there are many different educational technology communities that use different names for the field, including, but not limited to, "technology enhanced learning", "e-learning", "instructional design", and "technology-based learning".

Czerniewicz (2010) argues that one of the factors for this structural problem of educational technology is its weak terminology, which brought about several specialist "languages" such as instructional design and computer-supported collaborative learning. Different disciplines add to the structural problem of educational technology: it is not vertically integrated. Ely (2008) emphasizes that the fragmentation of educational technology also stems from different educational systems around the world that are shaped by local culture and the possibilities of the infrastructure. As a result, there are many disjoint scientific communities in educational technology (Gillet et al. 2009), a phenomenon also found in other interdisciplinary fields - e.g. in Human-Computer Interaction (Henry et al. 2007).

Given the fragmentation of the educational technology and the steady increase of scientific literature, it can be very cumbersome to keep an overview of the field. Knowledge domain visualization (KDViz) is a possible remedy to this problem. It aims at the representation of knowledge by "visually painting a picture of the scientific development and evolution of a domain" (Faisal et al. 2006). Fisichella et al. (2010) even argue that mappings of the field might help to overcome the fragmentation in educational technology by building awareness among researchers of the different sub-communities.

Knowledge domain visualizations are usually based on citations. Small (1973) and Marshakova (1973) proposed co-citation as a measure of subject similarity and co-occurrence of ideas. This relationship can be used to cluster documents from a field and map them in a two-dimensional space (see Figure 1 for a simplified example). Co-citation is an empirically well validated measure that has been applied to many fields, including educational technology. Furthermore, co-citation provides more stable results over time than content-based measures such as co-word analysis (Leydesdorff 1997). There is, however, a significant problem with citations: they take a long time to appear. Therefore, co-citation visualizations have to deal with a serious time lag.

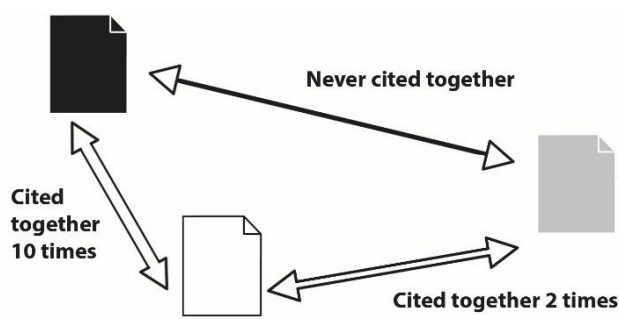

Figure 1: Knowledge domain visualization based on co-citations. Documents that are often cited together are placed closer to each other than those that are seldom or never cited together.

In comparison to traditional publications, social media offers a much faster means of communication. Services like Twitter and ResarchGate have been adopted by a large number of scientists in the last few years (Van Noorden 2014). In the wake of these developments, measures derived from these alternative communication channels (in short "altmetrics") have been discussed as a complementary source for the quantitative analysis of scientific output (Priem et al. 2010). Altmetrics are earlier available, potentially shortly after the publication of an article.

Social reference management systems in particular, have enjoyed a high popularity among educational technology researchers and the general scientific population (Kraker & Lindstaedt 2011, Kraker et al. 2012). Systems like Bibsonomy, CiteULike and Mendeley enable researchers to manage their references in a personal library and to share them with other users. In addition, readership numbers give an indication how many users have added a publication to their libraries, making it possible to see the literature "through the eyes of the readers" (Rowlands and Nicholas 2007). There are first indications that co-readership may serve as a measure of subject similarity (Jiang et al.

2011, Kraker et al. 2015). The topical relationship established by co-readership can be exploited for visualizations in the same manner as co-citation (see Figure 2), potentially leading to more recent representations of the field.

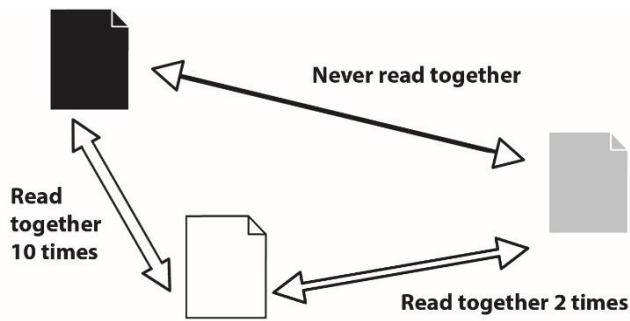

Figure 2: Knowledge domain visualization based on readership. Documents that are often read together are placed closer to each other than those that are seldom or never read together.

In this paper, I present the evaluation of a knowledge domain visualization of educational technology that is based on co-readership in the online reference management system Mendeley. The visualization has been evaluated by (1) a qualitative comparison to other visualizations of the field and by (2) expert interviews with researchers from the field to validate the results from the first evaluation.

## 2  Knowledge Domain Visualization of Educational Technology

The knowledge domain visualization of educational technology evaluated in this paper can be seen in Figure 3. Topic areas are represented as blue bubbles. Documents attached to each area are placed inside the circles. The metadata of the document is shown in the document representation itself. It consists of the most common metadata: title, author(s), year, and where it was published. The size of the circle represents the combined readership of documents in the area, whereas the size of the document signifies the number of readers it has.

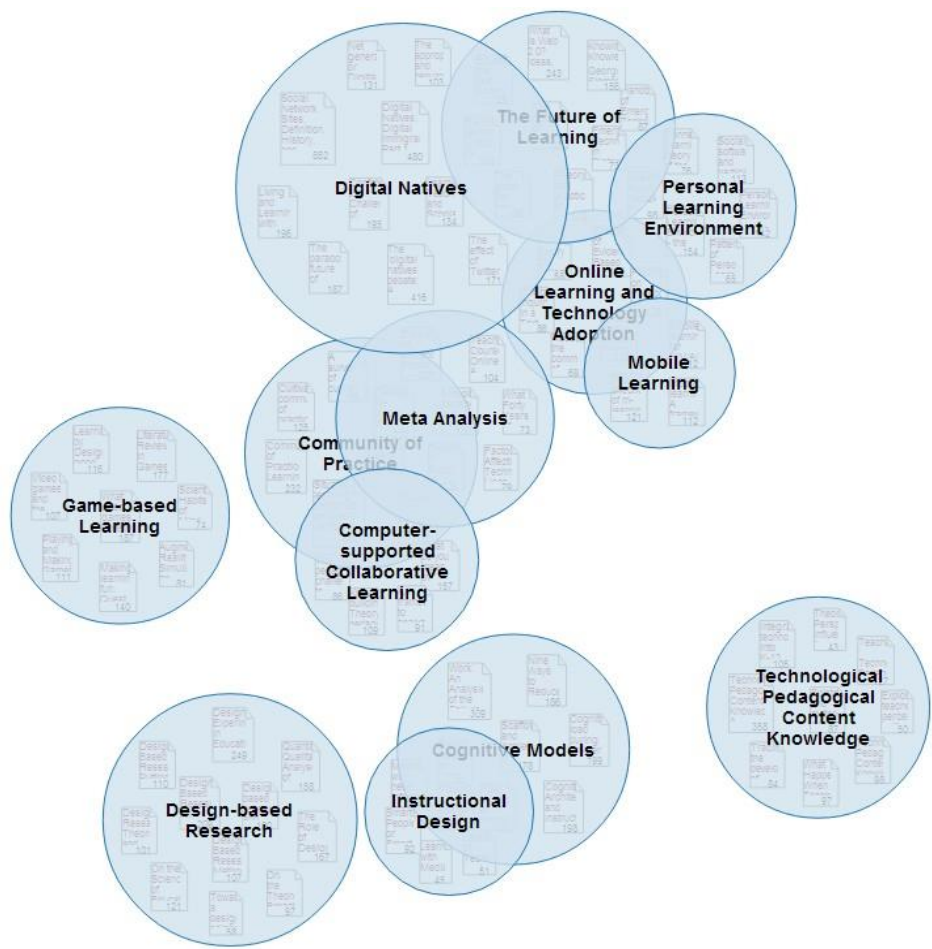

Figure 3: Knowledge domain visualization of educational technology. The blue bubbles represent topic areas. Documents attached to each area are placed inside the circles. The size of the circle represents the combined readership of documents in the area.

Users can interact with the visualization. In that regard, the well-tested approach of "overview first, zoom and filter, then details-on-demand" (Shneiderman 1996) was followed. The interface was designed in such a way that most of the exploration can be done within a single window. Once a user clicks on a bubble, he or she is presented with relevant documents for that topic area (see Figure 4). The dropdown on the right displays the same data in list form with an added abstract. By clicking on one of the documents, a user can access all metadata for that document. If a preview is available, one can retrieve it by clicking on the thumbnail in the metadata panel. In addition, a user can filter the publications by entering terms in the search field on top of the list. Only publications that contain all of the search terms (Boolean AND) are displayed within the bubbles and the list. The list can be sorted by title, area, and number of readers to facilitate exploration via the list format. The visualization prototype can be accessed at http://labs.mendeley.com/headstart; the source code has been published as well (Kraker & Weißensteiner 2012).

The visualization was created using readership data from the online reference management system Mendeley. Among other things, Mendeley enables users to maintain a reference library and share references with other users. Mendeley keeps track of article readership and augments each article with the number of people that have added this article to their library (Henning and Reichelt 2008). The data set used for the visualization was composed on 10 August 2012.

The most read papers of Mendeley users from educational technology served as the basis of the visualization. A threshold of 16 readers was introduced as selection criterion. This means, a document needs to have been added to at least 16 libraries owned by users who identified themselves as being in the field of educational technology to be included in the analysis, leading to a total of 91 documents. This threshold was introduced to cancel out noise in the data, and to present users with a manageable amount of documents. For these 91 documents, the co-readership matrix was determined. On top of that matrix, multidimensional scaling, hierarchical clustering, and force-directed placement were performed. For the naming of the areas, a heuristic was used that produces suggestions based on text mining services OpenCalais and Zemanta. All details of the technical implementation can be found in Kraker et al. (2015).

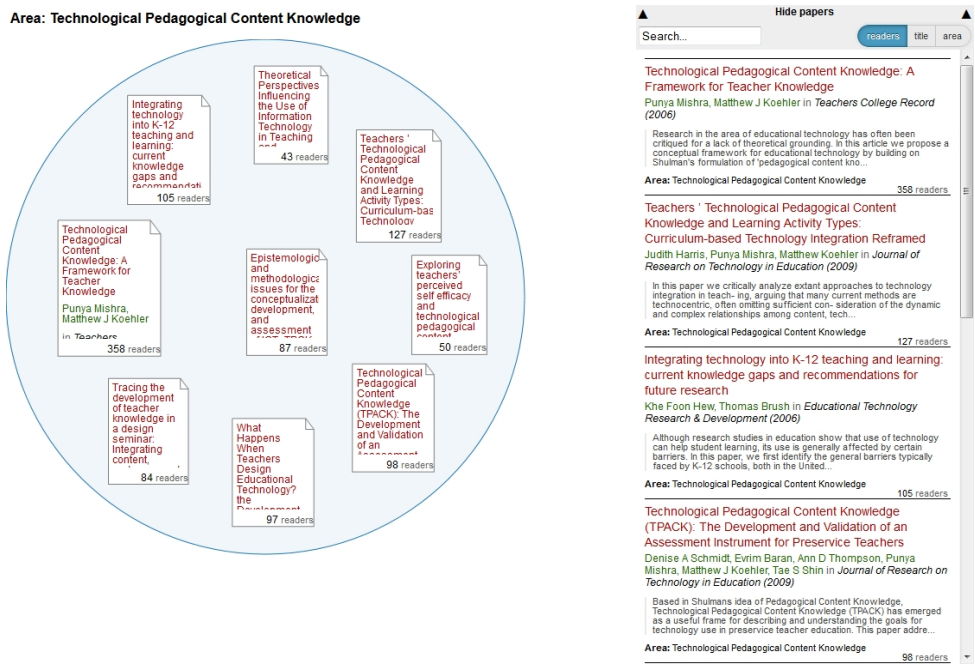

Figure 4: Topic area of Technological Pedagogical Content Knowledge. The metadata of the document is shown in the document representation itself. It consists of the most common metadata: title, author(s), year, and where it was published. The size of the document signifies the number of readers it has.

## 2.1 Literature Base

There are five different types among the publications contained in the visualization. The majority are journal articles (71 items, or 78%), followed by reports (7), books (6) and book chapters (5), and conference papers (2). The 71 journal articles were published in a variety of journals. The highest number of articles were published in Computers & Education (8), followed by Educational Technology Research & Development and The Internet and Higher Education (both 6) and Review of Educational Research, Educational Researcher and Educational Psychologist (all 5). These journals are among the most highly regarded journals in the field.

80% of publications were published from 2003 onwards, meaning that they were younger than ten years at the time of data collection (10 August 2012). Most documents were published in 2009 (see Figure 5). The median age of publications is 6.0 years (Mean = 7.3 years). Classics within the field are still contained in this visualization; for the most part they inform research that is still prevalent today. Examples are "Situated learning: Legitimate peripheral action" (Lave and Wenger 1991) or "Cognitive load during problem solving: Effects on learning" (Sweller 1988). An exception is the area "Instructional Design" which contains only documents that were published before 2003. Here, the classic media debate between Clark and Kozma is represented, as well as other older papers relating to instructional design. Even though the literature is comparatively new (the highest number of documents was published in 2009), there is a sharp decline in the years 2010 and 2011.

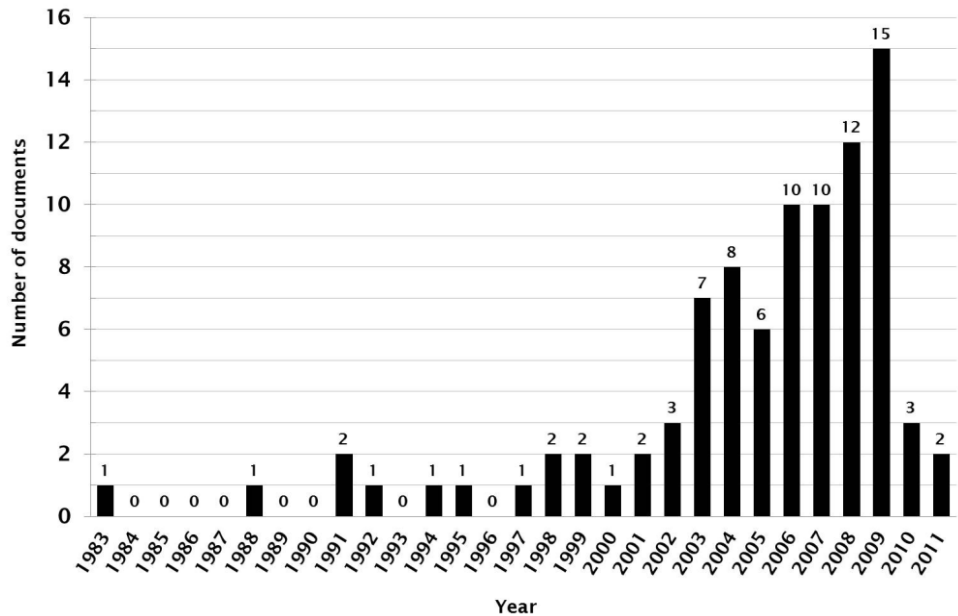

Figure 5: Distribution of publication years of documents in the visualization (n=91 documents)

*2.2 Topical Representation*

Cluster analysis of the co-readership matrix revealed 13 areas of educational technology research with a combined readership of 13,630. In Table 1, an overview of the areas can be seen. The areas differ in terms of the number of documents and the number of readers. Digital Natives has the highest readership with over 20% of all readers. The area received its name from the documents that revolve around the digital natives' debate, started by Prensky (2001). Several papers in the cluster discuss what kinds of skills are needed in the digital world as well as the role of new literacies such as media literacy and information literacy needed for a participatory culture. Digital Natives has twice the readership of the second largest area: Design-based Research (DBR), a research methodology of educational interventions. DBR sports the most documents (11) of all areas. Several founding publications, e.g. by Collins (1992) are represented in this area.

Six areas are of medium size (8.7-7.3% of combined readership). The Future of Learning is the largest of these areas. It contains publications describing technological developments, changes in knowledge, and the future of educational institutions in the digital age, but also publications complementary to the areas Online Learning, Digital Natives and Computer-supported Collaborative Learning. The second largest cluster in this group, Community of Practice, covers four highly read publications, including the citation classic "Situated learning: Legitimate peripheral participation"' by Lave and Wenger (1991). The area Cognitive Models contains three papers dealing with cognitive load theory first described by Sweller (1988). Two other papers relate to cognitive load as well, but deal with cognitive architecture in general. They represent a debate between Kirschner (2006) and Hmelo-Silver (2007).

Also part of the mid-sized areas, Technological Pedagogical Content Knowledge (TPACK) and Game-based Learning are two very coherent areas. TPACK is a conceptual framework for educational technology, which asserts that teachers need technological, pedagogical and content knowledge, as well as an understanding of the interplay of these three components (Koehler and Mishra 2005). More than half of the articles in this area sport the two original proponents of the framework, Koehler and Mishra as co-authors. Game-based Learning deals with educational games. Gee is the dominant author in this area with three papers, see e.g. Gee (2003). Related topics covered are virtual words and augmented reality.

Table 1: Overview of areas in the knowledge domain visualization

| **Area** | **No. Documents** | **No. Readers** | **% Readership** |
|---|---|---|---|
| **Digital Natives** | 10 | 2865 | 21.0% |
| **Design-based Research** | 11 | 1477 | 10.8% |
| **The Future of Learning** | 9 | 1183 | 8.7% |
| **Community of Practice** | 4 | 1175 | 8.6% |
| **Cognitive Models** | 6 | 1169 | 8.6% |
| **Technological Pedagogical Content Knowledge** | 9 | 1049 | 7.7% |
| **Game-based Learning** | 8 | 993 | 7.3% |
| **Meta Analysis** | 8 | 991 | 7.3% |
| **Personal Learning Environment** | 6 | 648 | 4.8% |
| **Online Learning and Technology Adoption** | 6 | 637 | 4.7% |
| **Computer-supported Collaborative Learning** | 5 | 615 | 4.5% |
| **Instructional Design** | 6 | 483 | 3.5% |
| **Mobile Learning** | 3 | 345 | 2.5% |
| Sum | 91 | 13630 | 100.0% |

Meta Analysis is the last of the mid-sized areas. It does not represent a coherent topical area but rather a collection of reviews/state-of-the-art analyses of the field. They encompass general educational technology research and theories, mobile learning, online learning, CSCL, and formative assessment.

Personal Learning Environment (PLE) is the largest of an array of smaller areas (under 5% of total readership). PLEs usually comprise a number of resources that are relevant to the learner in an online learning system based on Web 2.0 services and social media components (Atwell 2006). Not all papers in this cluster deal with PLEs directly; other topics include social software, participatory learning, and connectivism. The next smaller area, Online learning and Technology Adoption is a very diverse area. It contains papers on online learning and related frameworks, but it also deals with technology adoption. Computer-supported Collaborative Learning (CSCL) is a well-established area within educational technology. The cluster contains just five papers, among them, however, are important authors in the field, e.g. Laurillard (2009) and Dillenbourg (1999).

Instructional Design contains among other papers the classic media debate between Kozma (1991) and Clark (1983), on whether the medium (or only the content) can influence learning. The area with the least readers and the least number of documents is Mobile Learning with just 3 documents and a combined readership of 345. All of them are related to learning with handheld devices.

The topic areas can be assigned to some rough meta-areas. On the top of the map (see Figure 3), social and technological developments are being discussed in Digital Natives and The Future of Learning. Beneath this meta-area, there is a large cluster of learning methods and technologies, spanning Mobile Learning, Personal Learning Environment, Online Learning and Technology Adoption, Community of Practice, and Game-based Learning. On the lower end, there is a cluster of areas that form the psychological, pedagogical, and methodological foundations of the field; the areas Computer-supported Collaborative Learning, Instructional Design and Cognitive Models relate to psychology, while Technological Pedagogical Content Knowledge relates to pedagogy. Research methods are represented by Design-based Research.

Another characteristic of knowledge domain visualizations based on multidimensional scaling, such as the one under evaluation, is that areas more central to the field are also more central in the visualization (Mccain 1990). Right in the center, the area Meta Analysis contains reviews of the field. Other central areas are CSCL, Communities of Practice, and Online Learning and Technology Adoption. The remaining areas are more peripheral to their respective meta areas. Game-based Learning, Design-based Research, and Technological Pedagogical Content Knowledge in particular are placed further to the edges of the visualization.

## 3 Evaluation I: Comparison to Knowledge Domain Visualizations of Educational Technology

A number of articles have presented visual overviews of educational technology in the past. In this section, I present an evaluation based on a qualitative literature comparison of the knowledge domain visualization based on readership patterns presented in the last section (hereafter: co-readership visualization) to two recent studies of the field by Cho (2012) and Chen and Lien (2011). Both studies employ co-citation analysis; see Table 2 for an overview of the two papers.

For the qualitative comparison, areas and topics related to these areas found in each of the studies were entered into a spreadsheet. I then compared this list to the topics found in the co-readership visualization and noted overlaps and differences in the coverage. Furthermore, I compared the spatial aspects of the visualization to the representation provided in both papers. The results of this comparison are reported in the following sections.

Table 2: Papers analyzed in the qualitative comparison

| Publication | Database | Method |
|---|---|---|
| Cho et al. (2012): "The landscape of educational technology viewed from the ETR&D journal" | Educational Technology, Research & Development 1989-2011 (800 articles) | Document co-citation analysis of 28 articles |
| Chen and Lien (2011): "Using author co-citation analysis to examine the intellectual structure of e-learning: A MIS perspective" | 127 articles from 27 journals, 379 Taiwanese dissertations | Author co-citation analysis of 40 authors |

### 3.1 Comparison to Cho et al. (2012)

In Cho et al. (2012), the authors carry out a document co-citation analysis based on 28 highly co-cited articles from a database of 800 ETR&D articles. The analysis results in a map with three distinctive clusters: "Media debate", "Learner control", and "Learning environments". It is important to note that the corpus of the co-readership analysis and that of the co-citation analysis is different in terms of the recency of the publications. In case of Cho et al. (2012), the mean age is 14.1 years, which is almost double the mean age in the co-readership visualization (7.3 years).

*Educational Technology as Seen Through the Eyes of the Readers*

The largest and most central cluster in the co-citation analysis of Cho et al. (2012) is "Learning environments". The size of this cluster is not surprising; some scholars even talk about an age of learning environments in educational technology in the 2000s (Mihalca and Miclea 2007). In the co-readership visualization, however, this concept has been split up into different areas. Learning environments are represented in three clusters in the co-readership visualization: (1) Personal Learning Environment, which is a learner-controlled aggregation of resources relevant for learning, (2) Design-based Research, which "blends empirical educational research with the theory-driven design of learning environments" (Design-based Research Collective 2003), and (3) Game-based Learning, which deals among other topics - with game-based learning environments. This fragmentation indicates that the notion of learning environments is not as important as it once was, but that is still being discussed within more recent concepts.

"Media debate" is a small cluster found on the lower left edge of the visualization. It only contains two papers, which represent the media debate between Clark (1994) and Kozma (1991). These two papers are also included in the co-readership visualization as part of the Instructional Design area. Just like "Media debate", Instructional Design is a relatively small cluster.

"Learner control", the third area in the visualization by Cho (2012), is also a smaller area found on the lower right edge of the visualization. Similar to "Learning environments", "Learner control" is not an area per se in the co-readership visualization, but learner control issues are discussed in other areas, primarily those that deal with participatory culture and online learning: Digital Natives, The Future of Learning, Online Learning and Technology Acceptance, and Personal Learning Environment. The emphasis on participatory culture in the co-readership visualization indicates that the discussion has changed from "allowing learners to choose the amount of practice, feedback, and review as they desire" (Schnackenberg and Sullivan 2000) to "the ability to create, to share ideas, to join groups, to publish" (Attwell 2007) and "new media literacies: a set of cultural competencies and social skills that young people need in the new media landscape" (Jenkins 2009). Furthermore, "Learner control" is a rather small area, whereas the corresponding areas in the co-readership visualization are among the largest in the visualization. This hints at the fact that learner control has evolved into a very important concept within educational technology.

There are certain commonalities with regards to area placement between the co-citation visualization by Cho et al. (2012) and the co-readership visualization. In the co-citation visualization, the clusters "Media debate" and "Learner control" are found on the edges of the map; in the co-readership visualizations, the areas representing these two clusters (Instructional Design, Digital Natives, The Future of Learning, and Personal Learning Environment) are also placed further to the edges of the visualization. In addition, in both visualizations these two clusters are on opposing ends: in the co-readership visualization, Instructional Design (representing "Media debate") is found on the lower edge, whereas the clusters representing "Learner control" are found on the upper edge (note that maps created with multidimensional scaling can be rotated, so long as the relative distances between the objects stay the same).

*4.2 Comparison to Chen and Lien (2011)*

Chen and Lien (2011) map the field of educational technology on the basis of literature in Management Information Systems (MIS), including 127 articles from 27 journals. For the 40 most highly cited authors from this dataset, Chen and Lien perform an author co-citation analysis. They map co-citation pairs using multi-dimensional scaling and clustering. This procedure leads to the following three areas: "Adaptive web-based learning", "The usage of IT in learning activities", and "Psychological research for using IT in learning". The resulting map shows that "The usage of IT in learning activities" is the most central area. The area "Psychological research for using IT in learning" is closely related, but further to the right edge of the map. The area "Adaptive web-based learning" is found on the far left edge of the visualization.

According to Chen and Lien (2011), the area "The usage of IT in learning activities" contains the topics "Design of e-learning" and "Cooperative learning". These topics are covered in the co-readership visualization in the areas Design-based Research, Instructional Design, TPACK, CSCL, Community of Practice and Online Learning. "Psychological research for using IT in learning" consists of the topics "User behavior and acceptance" and "TAM and satisfaction" (represented in Online Learning and Technology Acceptance, and Digital Natives), as well as "Social cognition and self-efficacy" (represented in Cognitive Models). The area "Adaptive web-based learning", however, is not represented in the co-readership visualization.

While "Adaptive web-based learning" is not represented in the co-readership visualization, the spatial representation of the other two areas of the co-citation analysis by Chen and Lien (2011) provides a comparable alignment to the co-readership visualization. Clusters representing learning methods and technologies are central in both visualizations, whereas the psychological foundations are closely related but located towards the edges of the map. The general similarity in the area location and alignment is an interesting result, given the fact that there are little overlaps between the literature bases of the two visualizations.

### *3.3 Areas Only Present in the Co-Readership Visualization*

Of all analyses, the co-readership visualization contains the most topic areas (13); therefore a number of areas are not represented explicitly in any of the other analyses. Mobile learning, for example, is completely missing in the co-citation analyses. While social factors are included in the two co-citation visualizations, the digital natives debate is not explicitly covered. Learning environments are extensively covered, but personal learning environments are not mentioned as a newer development in this area. Communities of practice are not considered, with the exception of situated learning mentioned by Cho et al. (2012). Finally, there are areas related to the teacher's perspective of educational technology in the co-citation analyses discussed, but the framework of technological pedagogical content knowledge is not reflected in any of them.

## 4   Evaluation II: Expert Interviews

Evaluation I does not represent an external validation in the sense that all of the visualizations in consideration, including the co-readership visualization, were based on scientific literature. In order to test for external validity, I conducted semi-structured interviews involving the use of the system with experts from the domain of educational technology. One of the goals of this evaluation was to gain further insights into the representativeness and recency of the map. In addition, experts were asked to comment on size and naming of the clusters as well as the spatial distribution of the research areas. Finally, the experts also evaluated document selection and distribution.

### *5.1   Participants*

10 researchers from the field of educational technology participated in the evaluation. Among the interviewees were 3 professors, 3 senior researchers, 4 PhD students. 4 experts came from Austria; the other six were from all over Europe and Israel. The self-assessed attribution to a field (multiple answers allowed) can be seen in Figure 6. Educational technology was non-surprisingly the top answer (9) followed by Psychology & Cognitive Science which was named by 5 out of 10 experts. When asked how long they had been involved in educational technology research, 3 experts reported experience between 20 and 30 years, 3 between 9 and 12 years, and 4 experts 4 to 5 years.

*5.2 Method*

The interviews were held either face-to-face or - when that was not possible - online. Four interviews were held in German and six in English. A typical interview would take approximately an hour. In the interviews, I first asked the participants to fill out and sign a form which included personal details and a standard declaration of consent. Then I started the audio recording. The first questions were meant as a warm-up and consisted of the previous experience of the participants in educational technology. I then introduced the participants to the visualization and asked them to familiarize themselves with the functionality and the content. After a few minutes of personal exploration, I started to ask questions about the content and the usability of the visualization. Questions included area coverage, area naming, recency, paper coverage, area closeness, area centrality and document selection.

The interviews were later transcribed and qualitatively analyzed. The transcripts were coded with MAXQDA and RQDA. The coding scheme consisted of the topics of interest mentioned above. If one of the topics appeared in the text, the appropriate passage was coded and paraphrased. These paraphrases were then refined into sub-categories in several iterations of qualitative interpretation.

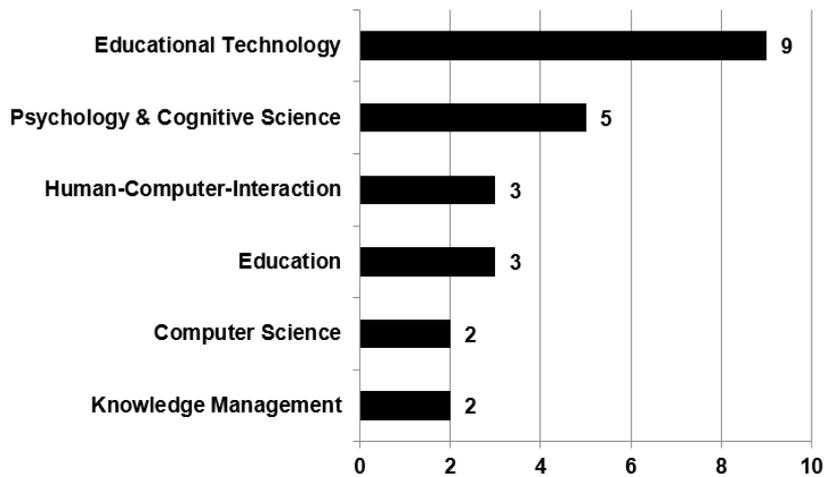

Figure 6: Discipline distribution among experts, multiple answers allowed (n=10 participants)

*5.3   Results*

*a)   Area Coverage, Naming, and Recency*

In relation to the area coverage, most experts agreed that the presented areas are part of educational technology. It was mentioned that these areas represented topics named in calls for papers and project proposals. Several experts commented that the areas give a representative overview of the field and that the topics and areas included are important to know for researchers familiarizing themselves with educational technology.

> "Currently these are important concepts which one needs to familiarize oneself when you start reading into the field." – P1

> "[...] the keywords I see […] all that seems to be reasonable. They correspond to what we see in the area, […]. It gives a view which is contemporary [...]" – P8

> "It reflects the field very well" – P10

The area selection, however, resonated better with experts with a focus on psychology and education than with experts from computer science. But even psychologists and educational scientists mentioned the lack of more computer science related areas such as learning objects, adaptive hypermedia, and intelligent tutoring systems. The overview was seen as being more education-oriented.

> "On first sight, all the areas seem to relate to psychology and pedagogy, and the all-technical areas are not readily identifiable. Learning Analytics would be one of those, a very hot topic. [...] Of course, you can write about something purely technological within the areas, but the label is apparently not." – P5

With regards to the area names, one participant mentioned that some of the areas are rather broad such as Instructional Design and Online Learning and Technology Adoption, whilst others are rather narrow such as Personal Learning Environments and Technological Pedagogical Content Knowledge (TPACK). Several experts were not familiar with TPACK, even though others saw it as an important and emerging area. A couple of experts mentioned that TPACK was one model within the wider field of teacher training in educational technology. They suggested renaming the area and including papers that relate to other models of teacher training.

> "[…] TPACK is kind of a really specific model, not even an area. So TPACK can be inside teachers' training, ICT in teachers' training. So maybe I would change it to teacher training, teacher knowledge." – P7

The Future of Learning on the other hand, struck participants as being too broad and also as an area that would be fluent over time. Digital Natives was a heavily debated area, not only regarding its name but also its size (see subsection "Area Size").

With respect to the recency of the visualization, several experts mentioned that the visualization represents a contemporary overview of the field. There was, however, also the notion that emerging fields and trends, such as learning analytics and massive open online courses (MOOCs) are not included in the overview. As a result, experts that are very concerned with the newest developments were less confident about the recency of the visualization and put it several years behind the current state-of-the-art. One participant summed up this conflict:

> "Very recent, yes. In my opinion, there are differences between computer scientists and psychologists, because we like to cite papers from the 60s, because the brain does not change and the patterns are the same. Computer scientists will look rather at the publication year and decide whether they are recent or not. But I think that concerning EU projects, these are the current topics, that are always named in calls; that's why I would say that they are recent topics." – P2

*b)   Paper Coverage*

Concerning paper coverage, the results were mixed. It was mentioned that paper selection works well for Design-based Research, CSCL, TPACK, Community of Practice, and Digital Natives. Regarding Mobile Learning and Game-based Learning, the experts stated that important papers and sub-topics were missing.

The experts were torn regarding the paper coverage in Cognitive Models and Instructional Design: some said that the papers work well, whereas others missed certain aspects. But it was not only the varying quality and coverage of papers that was being discussed in this context, but also their age. As the experts noted, certain areas contain rather new literature whereas others seem to consist of more fundamental work. This was troubling for computer scientists in particular.

*c)   Subject Closeness*

When asked, whether the ordination of the areas represented their view of the field in terms of subject similarity, the experts' answers where mostly positive. CSCL and Communities of Practice were often named as working well together. The same goes for Instructional Design and Cognitive Models. Digital Natives and the Future of Learning were also mentioned as fitting matches, as they form a cluster of areas of social and technological innovation. Participants also identified a technology-oriented cluster, including the areas PLEs, Mobile Learning, and Online Learning with the possibility of adding Game-based Learning. One participant made an interesting distinction between a collaborative cluster and a cognitive cluster:

> "Traditionally, there are cognitively-oriented layers versus collaboratively-oriented layers. These are the cognitive psychological approaches [Design-based Research, ID, Cognitive Models], whereas these are the collaborative ones [CSCL, Communities of Practice]. Here are different environments [PLE, Mobile Learning, Online Learning], and this is more about the future, maybe rather sociological approaches [Digital Natives, Future of Learning]. That works quite well." – P10

Some experts were critical about the ordination, especially those who missed the computer science based areas in the visualization. Several participants mentioned that Game-based Learning should be closer to Digital Natives as those two areas are often combined. The relative isolation of Game-based Learning was generally not well received. It was mentioned that this area should be closer to Mobile Learning, and that both should overlap with CSCL. Furthermore, TPACK was also seen as too separated.

> "In our papers, we often argued that Game-based Learning is especially suitable for young people, because they are digital natives and engage with technology in their daily life. That is why I would put them closer together." – P5

*d)   Area Size*

Area size was a controversial topic in the interviews. For many experts, the area Digital Natives was too large. They noted that this does not correspond with their perception of the importance of the area in the field. Some noted though that Digital Natives might be the most interesting area for researchers from other fields, resulting in higher readership numbers.

> "Digital Natives is kind of an odd black sheep, I did not expect to see it here. Also it is huge and it is really really general." – P7

*e)   Centrality*

As mentioned above, another characteristic of knowledge domain visualizations that are created with MDS is that areas, which are more central to the field are in the center of the visualization. When I asked the experts about this, I got many conflicting statements which hint at a high level of disagreement with regards to the centrality of the field. For some, the areas in the center made sense, as they were either important to the field (such as CSCL and Community of Practices), or – in the case of Meta Analysis – as they were topically spread over other clusters.

> "From my point of view, this [CSCL] is the center, so yeah, I agree, sure. [...] For me, CSCL can include a lot of research around." – P4

> "Meta Analysis in the center, that would make sense, because this is surely something that many readers from different backgrounds would follow." – P10

Others were not so sure about Meta Analysis being in the center, and some also questioned CSCL and Community of Practice.

> "[…] well of course the papers may be more important because they are meta analysis but it is not a topic which is more important, more central than others. So, maybe a bit misleading." – P6

Several participants would have liked to see more peripheral areas like Instructional Design and Design-based Research closer to the center, along with some of the more concrete learning technologies and methods. In general, for those experts who saw the difference between central versus peripheral areas as "specialized versus general communities" or "new versus emerging areas", centrality seemed to work better. For those that saw it as "influential versus non-influential areas", it seemed not to work as well.

## 5   Discussion and Conclusion

The evaluation showed that the presented knowledge domain visualization based on co-readership patterns is a recent representation of pedagogical and psychological research in educational technology. The qualitative comparison revealed that topics covered in more recent literature such as participatory learning and technological pedagogical content knowledge are better represented in the co-readership visualization than in comparable co-citation visualizations. The expert interviews continued the notion that the co-readership visualization is a recent representation, but they also revealed that some of the most recent developments such as MOOCs are not included.

The evaluations also showed that areas related to computer science such as adaptive hypermedia are not well-covered in the co-readership visualization. One possible explanation for this is that in Mendeley, educational technology is a sub-discipline of education. This sample bias in usage statistics was first mentioned by Bollen and van de Sompel (2008) and is also discussed in Kraker et al. (2015). A similar problem exists in citation studies, where the definition of the corpus usually imposes a bias, e.g. to a certain field as in Chen and Lien (2011). It is therefore crucial that the characteristics of the underlying user base of such visualizations are known; otherwise, it is not possible to draw informed conclusions.

The qualitative comparison showed that the co-readership analysis covers more areas than the co-citation analyses. There is still room for improvement though, as it was revealed that in some instances important papers were missing. In the future, it will be therefore important to include more papers in the visualization to have a better coverage of seminal publications. It may also be worthwhile to add further layers to the visualization, making it possible to drill down within a given sub-area.

An analysis of the spatial features of the maps in the qualitative comparison showed that there were many similarities among the maps created using co-citation and the co-readership visualization. The topical similarity also worked well, with only a few exceptions. Experts were torn, however, on the question of what the centrality of an area implies. The same is true for the size of the areas. Therefore, it will be important to conduct further research into the meaning of these concepts and provide users of the visualization with an adequate explanation.

In conclusion, readership-based maps seem promising, especially in dynamic and dispersed fields such as educational technology. Further research, however, is required to deal with the weaknesses revealed in the evaluation.

## Acknowledgements

I would like to extend sincere thanks to all participants in the expert interviews. I would also like to thank Sebastian Dennerlein, Dieter Theiler, and Dominik Kowald who assisted in transcribing and coding the interviews. Figures 1 and 2 were created by Maxi Schramm. The research presented in this work is in part funded by the European Commission as part of the FP7 Marie Curie IAPP project TEAM (grant no. 251514). The Know-Center is funded within the Austrian COMET program - Competence Centers for Excellent Technologies - under the auspices of the Austrian Federal Ministry of Transport, Innovation and Technology, the Austrian Federal Ministry of Economy, Family and Youth, and the State of Styria. COMET is managed by the Austrian Research Promotion Agency FFG.